\documentstyle[aps,prl]{revtex}  
\begin{document}
\draft

\twocolumn[\hsize\textwidth\columnwidth\hsize  
\csname @twocolumnfalse\endcsname              

\title{\bf Comment on ``Lattice QCD analysis of the  \\
strangeness magnetic moment of the nucleon''} 

\author { Chun Wa Wong}

\address{
Department of Physics and Astronomy, University of California, 
Los Angeles, CA 90095-1547}

\date{4 September 2002}

\maketitle

\begin{abstract}
The recent chirally extrapolated result of Leinweber and Thomas 
\protect{[Phys. Rev. D  {\bf 62}, 074505 (2000), or LT]} for the 
nucleon strangeness form factor  $G_M^s(0)$ = -0.16 $\pm$ 
0.18 $\mu_N$ differs markedly from the 
earlier result -0.75 $\pm$ 0.30 $\mu_N$ obtained by Leinweber 
\protect{[Phys. Rev. D {\bf 53}, 5115 (1996)]} from the 
same lattice data. An unresolved problem in the LT analysis of 
lattice data is identified and addressed. A value of 
$G_M^s(0)$ = -0.55 $\pm$ 0.37 $\mu_N$ is obtained at 
$R^s_d$ = 0.55 by extrapolating only the nucleon isoscalar 
lattice data.
\end{abstract}


   ]  

\narrowtext

Past lattice calculations made by Leinweber and collaborators 
\cite{LWD91,Le92} of baryon magnetic moments (MMs) have 
been restricted to massive quarks equivalent to pion masses 
$m$ greater than 600 MeV. The calculated moments have to 
be extrapolated to the physical pion mass $m_\pi$ = 140 MeV 
before comparison with experiment. Extrapolating 
linearly in $m^2$, Leinweber \cite{Le96} has estimated the 
nucleon strangeness form factor at momentum 
transfer $Q^2=0$ to be  $G_M^s(0)$ = -0.75(30) in units of 
$\mu_N$. (The number inside parentheses gives the statistical 
uncertainty in the last digits.)
A more sophisticated nonlinear extrapolation of the same 
lattice data using a chiral extrapolation ($\chi$E) formula 
proposed by Leinweber, Lu and Thomas (LLT) \cite{LLT99} 
has recently been made by Leinweber and Thomas 
(LT) \cite{LT00}. The result is -0.16(18). 

This LT estimate contains an unresolved problem caused by 
the use of an extrapolated valence $u$-quark contribution 
$u_{\Xi^0}$ for the baryon $\Xi^0$ that 
disagrees with experiment to renormalize another extrapolated 
term $u_n$. The purpose of this Comment is to suggest that 
the lattice data should be renormalized before the $\chi$E, 
and to point out that $G_M^s(0)$ can be extracted more 
readily from the nucleon isoscalar lattice data. The result is 
-0.55(37), where the error comes from uncertainties in the 
extrapolation formula, in the least-square fitting, and in the 
SU(6) constants $F$ and $D$. 

To extrapolate a MM term $\mu(m)$ calculated on a 
lattice for several large values of the pion mass $m$ to its
physical value at  $m_\pi$, LT use a Pad\'e 
approximant given by their Eq. (13) (denoted here LT13) that 
is the inverse of a quadratic Taylor approximant in $m$.
The extrapolated results, needed in this Comment, 
are read from Figs. 4 and 5 of \cite{LT01} and given in 
Table \ref{table1}. They agree well with experimental values
\cite{PDG00} where known, as pointed out in \cite{LLT99} for 
$\mu_p=p$ and $\mu_n=n$. 

For technical reasons, LT use a simple quadratic Taylor 
approximant in $m$, Eq. (14) in LT and called here LT14, for 
$u_{\Xi^0}$. (In terms of $s$ = $m^2$, LT14 
is the often-used linear approximant improved by a certain 
leading nonanalytic (LNA) term $\chi m$.) Table \ref{table1}
shows that the extrapolated value for $u_{\Xi^0}$ differs 
significantly from the experimental value $\Xi^0 - \Xi^-$. 
A concise review of these results in greater details has been 
given by \cite{LT01}. 

The discrepancy from experiment is not due to the use of 
LT14, as I have explicitly obtained an LT13 fit 
giving the same extrapolated value of $u_{\Xi^0}$ = -0.37 
(and -0.38 for LT14).  
The discrepancy comes instead from the sign of $\chi$ that
controls the curvature direction caused by the LNA term. 
In the LT model, $\chi$ has opposite signs for $u_n$ and 
$u_{\Xi^0}$, because the dominant pion cloud has opposite 
signs, namely $n \rightarrow$ $p\pi^- \rightarrow n$ 
compared to $\Xi^0$ $\rightarrow$ $\Xi^-\pi^+$ 
$\rightarrow \Xi^0$ \cite{LT00}. This sign makes it difficult 
for the $\chi$E of $u_{\Xi^0}$ to curve towards the  
experimental point. This is the unresolved problem in LT 
mentioned at the beginning of this Comment

To extract $G_M^s$ (dropping the $Q^2$ argument for 
notational simplicity), LT use a certain ``ratio'' method of 
Leinweber \cite{Le96} under which the disconnected-loop 
(DL) contribution $\mu_N^{\rm DL}$ of the nucleon sea 
is calculated from one of the following formulas:

\begin{eqnarray}
\mu_N^{\rm DL} ={1\over 3} (p + 2n - 
f_nu_n), \quad f_n =  {\Xi^0 - \Xi^-\over u_{\Xi^0}};
\label{Leinweber1}
\end{eqnarray}

\begin{eqnarray}
\mu_N^{\rm DL} ={1\over 3} (2p + n - 
f_pu_p), \quad f_p =  {\Sigma^+ - \Sigma^-\over u_{\Sigma^+}}.
\label{Leinweber2}
\end{eqnarray}
That is, the extrapolated value of $u_n$ is not used directly, 
but only after renormalization by $f_n$. Whenever $u_{\Xi^0}$ 
extrapolates correctly to its experimental value $\Xi^0 - \Xi^-$, 
the correction factor $f_n$ is 1 and therefore has no effect. The 
correction factor differs from 1 only when the extrapolation is 
unsuccessful. Then the argument for using the nontrivial 
correction factor $f_n \not= 1$ is that the ratio $u_n/ u_{\Xi^0}$ 
that appears has smaller systematic errors \cite{Le96}. 

The large renormalization $f_n$ = 1.62 found for Eq. 
(\ref{Leinweber1}) is the direct consequence of the 
aformentioned failure of the LT extrapolation for $u_{\Xi^0}$. 
In contrast,  $f_p$ = 0.973 needed in Eq. (\ref{Leinweber2}) 
contains little renormalization because LT13 works well there. 
The LT renormalized values (actually $f_iu_i$) are shown 
in Table \ref{table1} in the column marked LTR with errors 
from LT.

Is this postulated cancellation of systematic errors {\it after}
 LT extrapolations real? On the lattice, $u_n$ and 
$u_{\Xi^0}$ have very similar values, and their ratios are 
likely to contain smaller systematic errors. Unfortunately, 
these ratios cannot be used directly in the $\chi$Es, and the 
$u$'s have to be extrapolated separately for the lattice data 
used. Hence the correct procedure is to renormalize
one set of lattice data to fit an available experimental MM 
(thus ensuring that $f_i$ = 1 always), 
and to use the same renormalization on the second set of 
lattice data {\it before} the $\chi$E. 

This scheme can readily be executed for $u_{\Sigma^+}$ 
since the original extrapolation is already close to experiment. 
With the lattice data for different $m$ actually correlated, 
moving up and down together, one can simply take 
$\{u_i + x\Delta u_i\}$ as the corrected input with the 
unknown $x$ chosen to be -0.23 to make the extrapolated
value of $u_{\Sigma^+}$ fit $\Sigma^+ - \Sigma^-$. 
The same $x$ is then used to renormalize the lattice data for 
$u_p$. The LT13 extrapolated result, $u_p$ =  4.16,  
agrees with that from Eq. (\ref{Leinweber2}).

When the same procedure is applied to $u_{\Xi^0}$, it reaches 
a minimum of only -0.52, a little short of the experimental 
value. (The solution for the input $y\{u_i\}$ scaled by an 
adjustable factor $y$ reaches down to only -0.47.) There are 
LT14 solutions
($u_p$ = 2.34, $u_n$ = -1.54) that are very different from 
those of the Leinweber equations. Since LT14 also does 
very poorly for the nucleon MMs themselves, giving $p$ = 
3.14 and $n$ = -2.39, it appears to be unreliable.

The lattice data can also be renormalized by tilting and bending. 
This is conveniently realized by including the experimental 
point in an LT13 fit and using the resulting set of ratios 
$\{u_i({\rm output})/u_i({\rm input})\}$ to renormalize 
the second set of lattice data before its $\chi$E. The resulting 
renormalized lattice data (RLD) extrapolate to $u_n$ = 
-0.64(15), showing only a quarter of the renormalization effect 
from Eq. (\ref{Leinweber1}), while $u_p$ =  4.13(24) agrees 
well with Eq. (\ref{Leinweber2}). These results suggest that 
Eq. (\ref{Leinweber2}) is reliable, but 
Eq. (\ref{Leinweber1}) greatly over-estimates the 
renormalization. The statistical errors shown are obtained by 
treating the lattice data as correlated, i.e., calculated from 
$\{u_i \pm p\Delta u_i\}$, where $p$ = $\sqrt{1-(N_p/N_d)}$ 
= $1/\sqrt{3}$, where $N_d$ = 3 is the number of lattice data 
and $N_p$ = 2 is the number of fitted parameters \cite{Le02}.

An alternative explanation of the discrepancy in 
$u_{\Xi^0}$ that cannot be excluded at the present time is 
that its $\chi$ parameter is different from that 
calculated in LT. For example, the $\chi$ parameter needed 
to fit the experimental point together with the original lattice 
data is $\chi(u_{\Xi^0})$ = 0.2 (-0.3/+0.5), 
showing the change of sign needed for the $\chi$E to curve 
towards the experimental point. In other words,  a relatively 
small change in the LT value of  $\chi(u_{\Xi^0})$ = -0.4 is 
enough to change its sign. In the rest of this Comment, I shall 
work only with the LT model.

The value of $G_M^s$ = -0.16(18) reported by LT is obtained 
from Eq.(\ref{Leinweber1}) using the ratio $u_n/u_{\Xi^0}$ 
= 1.51(37). (The reported statistical error of 0.18 for 
$G_M^s$ is a misprint. It should read 0.27.)  LT has also 
obtained $G_M^s$ = -0.57(42) from 
Eq.(\ref{Leinweber2}) using the ratio 
$u_p/u_{\Sigma^+}$ = 1.14(11), unchanged  from 
the Leinweber linearly extrapolated value of 1.14(8). This is 
the better estimate because the renormalization involved is 
so much smaller and better established.

The second issue raised in this Comment is that the nucleon 
isoscalar quantity $G_M^s$ can be extracted from nucleon 
isoscalar lattice data alone without using any information from  
$u_{\Xi^0}$ or $u_{\Sigma^+}$. The resulting physical 
picture for $G_M^s$ is sufficiently simple to permit a rather 
unique answer to be obtained for any extrapolation formula 
used such as LT13. Rough estimates of other uncertainties 
can then be made readily.

The quantity to be extracted is the isoscalar DL contribution 
that, under isospin symmetry, is simply

\begin{eqnarray}
\mu_N^{\rm DL} & = & {1\over 2} (p + n) - 
{1\over 6}(u_p + u_n).
\label{isoscalar}
\end{eqnarray}
The numerical answer is $\mu_N^{\rm DL}$ 
= -0.18(5) using 
the experimental value of $p+n$ and the unrenormalized LT 
results shown in Table \ref{table1}. The answer is -0.10(7) 
with the original LT renormalization, and -0.14(9) using 
RLD to renormalize $u_p$ and $u_n$. The statistical errors 
shown are obtained from those in $u_p$ and $u_n$ treated as 
independent.

To do better in both value and error, it is necessary to 
extrapolate $u_p$ + $u_n$ together. Fortunately, there is 
enough information in the lattice data tabulated in LWD 
\cite{LWD91} to extract the 
covariance $\langle (\Delta u_p)(\Delta u_n) \rangle$ and 
to calculate the statistical error for any linear combination  
$au_1$ + $bu_2$ on the lattice. The combination is then 
extrapolated to $m_\pi$ by using the chiral parameter 
$a\chi_1$ + $b\chi_2$.

The least-square fitting program and input lattice data used 
here are first validated by checking against published results. 
The top panel of Table \ref{table1} gives a comparison 
with the results of LLT \cite{LLT99} for the nucleon MMs 
$p$ and $n$. The data used come from two independent 
sources \cite{LWD91,WDL92}. My errors are obtained by 
assuming that all lattice data are correlated (i.e., using
$N_p$ = 2 and $N_d$ = 6). The chiral parameters $\chi$ are 
defined in terms of the SU(6) 
constants $F$ and $D$ \cite{JLMS93,LT00}. The one-loop 
corrected values \cite{JLMS93} are used in both LLT and LT.
The table shows good agreement in both 
extrapolated values and fitting errors. 

In the second panel of the table, extrapolations are obtained 
for the correlated lattice data \cite{LWD91} used in LT. 
Both values and errors agree with their results.

The third panel of Table \ref{table1} shows the best results 
from Tabel \ref{table2} obtained by extrapolating $u_p$ 
$\pm$ $u_n$ together, using different $\chi$ parameters for 
flavor and valence contributions. The quality of this single-step 
extrapolation is first checked for the nucleon MMs. The 
isovector moment $p-n$ = $u_p$ - $u_n$ on the lattice 
because the isoscalar sea contributes nothing. Table 
\ref{table2} shows that the $p-n$ result of 4.65 for 
LT13 agrees well with the value of 4.68 from separate 
extrapolations. Both differ somewhat from the value 
$u_p$ - $u_n$ = 4.81 from separate extrapolations. 
The isoscalar moment extrapolates to $p+n$ = 1.03, 
in agreement with the value of 1.10 from separate 
extrapolations and with experiment. In contrast, LT14 
appears inadequate for both nucleon MMs.

The much more successful formula LT13 
proposed by LLT has the correct $1/m_q \propto 1/m^2$ 
behavior for heavy quarks. Its comparative success in chiral 
extrapolations comes from a reduction of the effect of the 
LNA term when placed in the denominator. Finally, 
the importance of having the correct heavy-quark limit is 
illustrated by using another inverse Taylor approximant M4D 
where the quadratic term in $m$ in the denominator is 
replaced by an incorrect quartic term. It works well too for 
the large isovector moment $(p-n)/2$ that dominates the 
nucleon MMs. This suggests once again that the LNA term 
should be in the denominator.

The extrapolated value of $u_p+u_n$ is also needed to
calculate $\mu_N^{\rm DL}$ from Eq. (\ref{isoscalar}).
It is obtained from the lattice data for $u_p$ + $u_n$ by 
extrapolating with a valence parameter $\chi^v$ different 
from the flavor value $\chi^f$ = 0 used for $p$ + $n$. 
The results obtained with different extrapolation formulas 
are shown in Table \ref{table2}. The very small statistical 
error found for $\mu_N^{\rm DL}$, denoted in the table as  
$\mu_N^{\rm DL}(1)$, comes from the fact 
that while each term in Eq. (\ref{isoscalar}) varies 
substantially as all lattice data move up and down together, 
the difference between the two terms remains essentially 
unchanged.

To determine if $p+n$ should be fitted in extracting 
$\mu_N^{\rm DL}$, I renormalize the lattice input data to 
$\{u_i + x\Delta u_i\}$ where $x$ is adjusted to reproduce 
the experimental value. The changes for all three formulas, 
also given in the table, are all quite small, showing that the 
procedure is quite robust. 

Unfortunately, there is a problem in the extrapolations for 
$p$ + $n$, and also for $p$ and $n$ separately. 
The lattice data used do not contain any DL contribution, 
which has in fact been assumed negligibly small in all LT 
extrapolations. Unpublished lattice data from Dong, Liu 
and Williams(DLW) \cite{DLW98,Liu02} seem to indicate 
that the DL contribution is not necessarily negligible in the 
mass range $m \approx$ 0.6-0.9 GeV of the LWD data. 
Unfortunately, the covariances for these DL lattice data 
have not been saved, thus precluding a quantitative estimate 
of the missing DL contribution for different $m$'s. This missing 
DL term is formally twice as strong and fractionally 4-6 times 
more important in $p$ + $n$ as it is in each MM separately. 
It gives an unknown systematic error to the extrapolated 
$p$ + $n$.

Given this unresolved problem in the lattice data, one might
want to use the experimental value of $p$ + $n$ in Eq. 
(\ref{isoscalar}). The results, shown in the table as 
$\mu_N^{\rm DL}(2)$, can differ significantly from
$\mu_N^{\rm DL}(1)$. The large error shown comes from 
the statistical fitting error of $u_p$ + $u_n$. 

These two estimates of $\mu_N^{\rm DL}$ agree if the 
extrapolated $u_p$ + $u_n$ agrees with experiment. It is 
therefore interesting to see what happens when the lattice data 
for $u_{\Xi^0}$ and $u_{\Sigma^+}$ are used to renormalize 
the lattice data for $u_n$ and $u_p$ {\it before} extrapolating 
$u_p$ $\pm$ $u_n$ by LT13 and M4D. The results, 
given in the lower panel of Table \ref{table2}, show that 
the extrapolated $p$ + $n$ by LT13 is now close to the 
experimental value. I take $\mu_N^{\rm DL}(2)$ as the 
better estimate because it contains no systematic error from 
$p$ + $n$ and a more conservative error estimate. This is 
the ``best'' result shown in Table \ref{table1}. The difference 
of 0.07 between LT13 and M4D shall be taken to be an 
additional uncertainty arising from incomplete 
knowledge of the correct $\chi$E formula. 
This large uncertainty has not been included in LT.

Errors also arise from the choice of the SU(6) constants 
$F$ and $D$ on which the $\chi$s depend. The $\chi$ 
parameter for the isoscalar MM $p+n$ is always $\chi^f$ 
= 0 for any $F$ and $D$. The valence parameter $\chi^v$ 
for $u_p+u_n$ is proportional to (and has the same sign as) 
\cite{LT00}

\begin{eqnarray}
\beta^v= -a(F+D)^2,  \quad a = {2\over 3}  
\left [{5 -6r + 9r^2\over (1+r)^2} \right],
\label{betav}
\end{eqnarray}
where $r \equiv F/D$. The factor $a$ is always positive, 
so that both $\mu_N^{\rm DL}$ and $G_M^s$ 
extracted from it are always negative in LT extrapolations. 
The factor $a$ has a minimum of 2.39 at $r$ = 0.74, and the 
values 2.40 and 2.45, respectively, at the one-loop 
corrected value of $r$ = 0.66 and the tree-level value of 
$r$ = 0.58 \cite{JLMS93}. Hence the error from uncertainties 
in $r$ is small and will be neglected.

With $r$ = 0.66, the uncertainty in $F+D$ can be determined
by treating it as a free parameter in LT13 fitted to both 
experimental and lattice data for each nucleon MM $p$ 
or $n$. The expanded data set used by LLT then gives

\begin{eqnarray}
(F+D)_p =  0.96(20),  \quad  (F+D)_n =  1.00(20), \nonumber \\
\Delta (u_p+u_n) =  0.21,  \quad  \Delta \mu_N^{\rm DL} =  0.04. 
\quad \quad
\end{eqnarray}
This yields a total error of 0.10 for $\mu_N^{\rm DL}$ 
reported in Table \ref{table1} when quadratically combined 
with the errors of  0.06 (from least-square fitting) and 0.07
(from the $\chi$E formula). 

The resulting values of  $G_M^s$ are shown in the lowest 
panel of Table \ref{table1} for  $R_d^s$ = 0.55 and 0.59 as 
used by LT and LLT, respectively. The value of -0.55(37) at 
$R_d^s$ = 0.55 agrees with the result -0.36(20) of 
DLW \cite{DLW98} obtained from a direct evaluation 
of the DL contribution. The value $R_d^s$ = 0.55 
used here is actually their ratio of directly evaluated strange 
to $u/d$ DL contributions.

From the perspective of Eq. (\ref{isoscalar}), the Leinweber 
Eq. (\ref{Leinweber1}) or (\ref{Leinweber2}) contains 
a spurious nucleon isovector term

\begin{eqnarray}
\Delta \mu_N^{\rm DL} = \mp{1\over 6} [p - n - 
f_i(u_p - u_n)],
\label{errormu}
\end{eqnarray}
where $f_i$ = $f_p$ or $f_n$. With $f_p$ close to 1, 
this term is very small in 
Eq. (\ref{Leinweber2}). This is another reason why Eq. 
(\ref{Leinweber2}) is better than Eq. (\ref{Leinweber1}). 

Additional systematic errors not taken into consideration 
here include effects from finite-volume and finite-spacing 
errors and from the quenched approximation not already 
simulated by the $\chi$E formula used \cite{LLT99}. They 
also include uncertainties in the ratio $R_d^s$ and 
differences between the LT model and 
true QCD, such as the contributions of neglected 
kaon loops and of $m_q{\rm ln}\,m_q$ terms, in as far as 
they affect the parameter analogous to  $\beta^v$ of 
Eq. (\ref{betav}) that drives the DL 
contribution in the $\chi$E. 

I would like to thank Drs. D. Leinweber and K.F. Liu for 
helpful comments.

\begin{table}
\caption{Comparison of extrapolated magnetic moment $\mu_i$ 
(in units of $\mu_N$) obtained by Leinweber, Lu and Thomas 
(LT) \protect{\cite{LLT99}} and by Leinweber and Thomas 
(LT) \protect{\cite{LT00}} without and with LT renormalization 
(LTR)  with experimental moments and with results obtained 
here using nucleon lattice data only (third panel of the table). 
The number inside parentheses gives the statistical uncertainty 
in the last digits. Estimates of $G_M^s(R^s_d)$ in the lowest 
panel are obtained from $\mu_N^{\rm DL}$ as defined by 
Eq. (\protect{\ref{isoscalar}}) using the experimental value 
of $p+n$.}
\begin{tabular}{ccccc}
$\mu_i$ & LLT/LT & Here  & LTR & Expt \cite{PDG00} \\  
\hline
$p = \mu_p$ & 2.85(22)  &  2.85(21)  &  & 2.7928...   \\
$n = \mu_n$ & -1.90(15) &  -1.91(15)   &  & -1.9130...  \\
\hline
$p$ &       & 2.89(19)       &        &    \\
$n$ &       &  -1.79(20)     &        &      \\
$u_{\Sigma^+}$ & 3.72(27)& 3.73(28)   &   & 3.618(27)  \\
$u_p$  & 4.26(25)  & 4.26(25)   &  4.12(40)   &             \\
$u_n$  & -0.56(15)  &-0.55(11)  & -0.90(22) &             \\
$u_{\Xi^0}$ &   -0.37(3)   & -0.37(3) &   & -0.599(14) \\
\hline
$u_p+u_n$ &   & 3.53(33)    &    &    \\
$p + n$ &      & 0.91(10)    &      & 0.8798...   \\
$p - n$ &      &  4.67(42)    &     & 4.7058...   \\
\hline
$\mu_N^{\rm DL}$ &  -0.18   & -0.15(10)  & -0.10 &   \\
$G_M^s(0.55)$ &  -0.65 &  -0.55(37) & -0.36 &  \\
$G_M^s(0.59)$ & -0.76  &  -0.65(43) & -0.42 &  \\
\end{tabular}
\label{table1}
\end{table}

\narrowtext
\begin{table}
\caption{Extrapolated nucleon magnetic moments obtained 
by using the chiral extrapolation formulas LT14, LT13 and 
M4D from nucleon lattice data without and with data 
renormalization to fit $p+n$. The lower panel gives results
obtained from renormalized lattice data (RLD) for $u_p$ 
and $u_n$ that use information from $u_{\Xi^0}$ 
and $u_{\Sigma^+}$.}  
\begin{tabular}{cccccc}
 & $p-n$ & $p+n$ & $u_p+u_n$ &  $\mu_N^{\rm DL}(1)$  & $\mu_N^{\rm DL}(2)$  \\  
\hline
Expt   & 4.7058... &  0.8798... &               &                 &   \\
LT14  & 5.50(23) &  0.77(5)   & 3.84(14) & -0.2560(2)  & -0.21(2)\\
&                &  0.88       &  4.18       & -0.2565  &   \\
LT13  & 4.65(40) &  1.03(13) & 3.89(41) & -0.136(5)  & -0.21(7)\\
&                &  0.88       &   3.42      & -0.130  &   \\
M4D  & 4.43(26) &  0.70(5)   & 3.24(16) & -0.189(5)  & -0.10(3)\\
&                &  0.88       &  3.87       & -0.206 &    \\
\hline
RLD:  &    &     &    &   & \\
LT13 & 4.67(42) & 0.91(10) & 3.53(33) & -0.134(4) & -0.15(6)\\
M4D & 4.44(26) & 0.67(4) & 3.12(15) & -0.186(4) & -0.08(3) \\
\end{tabular}
\label{table2}
\end{table}


\begin{thebibliography}{99}

\bibitem{LWD91}
D.B. Leinweber, R.M. Woloshyn, and T. Draper, Phys. Rev. D 
{\bf 43}, 1659 (1991).

\bibitem{Le92}
D.B. Leinweber, Phys. Rev. D {\bf 45}, 252 (1992).

\bibitem{Le96}
D.B. Leinweber, Phys. Rev. D {\bf 53}, 5115 (1996).

\bibitem{LLT99}
D.B. Leinweber, Ding H. Lu and A.W. Thomas, Phys. Rev. D {\bf 60}, 034014 (1999).

\bibitem{LT00}
D.B. Leinweber and A.W. Thomas, Phys. Rev. D {\bf 62}, 074505 (2000).

\bibitem{LT01}
D.B. Leinweber and A.W. Thomas, Nucl. Phys. A {\bf 684}, 35c (2001).

\bibitem{PDG00}
Particle Data Group, E. Phys. J. C {\bf 15}, 1 (2000).

\bibitem{Le02}
D. B. Leinweber, private communication.

\bibitem{WDL92}
W. Wilcox, T. Draper, and K.F. Liu, Phys. Rev. D {\bf 46}, 1109 (1992).

\bibitem{JLMS93}
Elizabeth Jenkins, Michael Luke, Aneesh V. Manohar and Martin Savage, Phys. Lett. B. {\bf 302}, 482 (1993).

\bibitem{DLW98}
S.J. Dong, K.F. Liu and A.G. Williams, Phys. Rev. D {\bf 58}, 074504 (1998).

\bibitem{Liu02}
K.F. Liu, private communication.

\end{thebibliography}
\end{document}